# Rigid Registration of Freehand 3D Ultrasound and CT-Scan Kidney Images

Antoine Leroy[1], Pierre Mozer[1,2], Yohan Payan[1], and Jocelyne Troccaz[1]

[1] Laboratoire TIMC-GMCAO, Faculté de Médecine, Domaine de la Merci,
38700 La Tronche, France
`Antoine.Leroy@imag.fr`
[2] Service d'Urologie et de Transplantation Rénale. CHU Pitié-Salpêtrière. AP-HP,
75013 Paris, France

**Abstract.** This paper presents a method to register a preoperative CT volume to a sparse set of intraoperative US slices. In the context of percutaneous renal puncture, the aim is to transfer a planning information to an intraoperative co-ordinate system. The spatial position of the US slices is measured by localizing a calibrated probe. Our method consists in optimizing a rigid 6 degree of freedom (DOF) transform by evaluating at each step the similarity between the set of US images and the CT volume. The images have been preprocessed in order to increase the relationship between CT and US pixels. Correlation Ratio turned out to be the most accurate and appropriate similarity measure to be used in a Powell-Brent minimization scheme. Results are compared to a standard rigid point-to-point registration involving segmentation, and discussed.

## 1  Introduction

Percutaneous Renal Puncture (PRP) is becoming a common surgical procedure, whose accuracy could benefit from computer assistance. The pre-operative imaging modality is CT, whereas either fluoroscopy or echography is used for intra-operative target visualization. A feasibility study on Computer-Assisted PRP has been carried out [3], in which the kidney surface, segmented from CT and localized US images, was registered using ICP. The study showed satisfying results; however it required a manual segmentation in both image modalities, which is not acceptable for a clinical use, especially intra-operatively.

We therefore investigated automatic CT/US registration. It was decided for the present study to propose and evaluate an automatic voxel-based registration algorithm, to avoid segmentation steps and to minimize user intervention.

Voxel-based registration methods have been deeply studied since 1995. Every method proposes a similarity measure and a cost minimization algorithm. Wells [12] first introduced Mutual Information (MI) combined with histogram windowing and a gradient descent algorithm. Maes [4] presented an interesting combination of MI and Powell-Brent (PB) search strategy. He also compared various search and multi-resolution strategies, and showed that PB was efficient with image subsampling [5]. Jenkinson [2], Studholme [11] and Sarrut [10] made a thorough comparison of different functional and statistical similarity measures.

Although those studies constitute the base of our research, none of them is applied to registering US images. We will therefore focus on the works of Roche [9], who registered 3D US of the brain with MRI, and Penney [6], who registered 2.5D US of the liver with MRI. The theoretical difficulty in registering CT and US is that the former gives information on tissues intensity, whereas the latter contains a speckled image of their boundaries. So a complex similarity measure and specific image preprocessing must be chosen.

This paper introduces a method to automatically register localized US images of the kidney onto a high-quality abdominal CT volume. The final algorithm uses image preprocessing in both modalities, Powell-Brent method as a search strategy, and Correlation Ratio (CR) as a similarity measure. Preliminary tests have been carried out on one data set, that allows us to draw the first conclusions on the method.

## 2 Image Preprocessing

### 2.1 CT Preprocessing

Penney [6] basically transformed the blurred MRI and US images of the liver into maps giving the probability of a pixel to be liver tissue or vessel lumen. However, as this process requires manual thresholding of the MRI, and as the segmentation of the kidney parenchyma is not a binary process, especially in the US images, we do not think that the technique can apply to our problem. Roche [9] proposed the combined use of the MRI image and its derivate, that we again decided not to use, because of the complexity of the bivariate correlation method and because of the observed chaotic correlation between CT gradient and US.

Our goal was to highlight the major boundaries in CT in order to increase the correlation with the US. The preprocessing of a CT slice consists in the superposition of a median blur and a bi-directional Sobel gradient from which we kept the largest connex components (fig. 1).

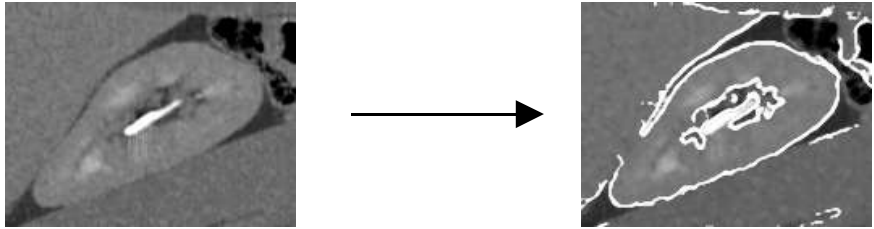

**Fig. 1.** Preprocessing of a CT oblique slice. The major boundaries are highlighted

### 2.2 US Preprocessing

**Speckle Removal.** US images are known to be low-quality gradient images, blurred by speckle noise. Still, the kidney, due to its small size and echogenic capsule, can be well and fully visualized through anterior access. The aim of preprocessing US images is to reduce the speckle, while preserving the boundaries of the organ. We applied the "sticks" filter proposed by Czerwinski [1], designed for that purpose.

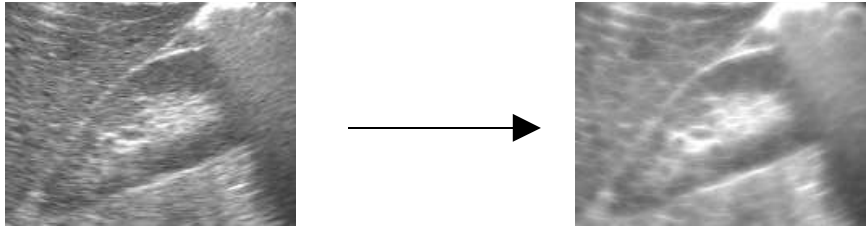

**Fig. 2.** Preprocessing of a US slice. Most speckle is removed while boundaries are preserved

**Shadow Removal.** Because they convey no relevant information for registration, acoustic shadows are removed in the US images, as proposed in [6]. The shadows are produced by large interfaces like the ribs or the colon that reflect quasi-totally the acoustic signal, the remaining waves decreasing in the distance in an exponential way. Shadow removal is based on the correlation between a US line profile and a heuristic exponential function. A shadow is detected when the correlation is higher than a given threshold, and when the maximum acoustic interface is close enough to the skin. Fig. 3 and 4 show the result of our automatic method on a sample slice.

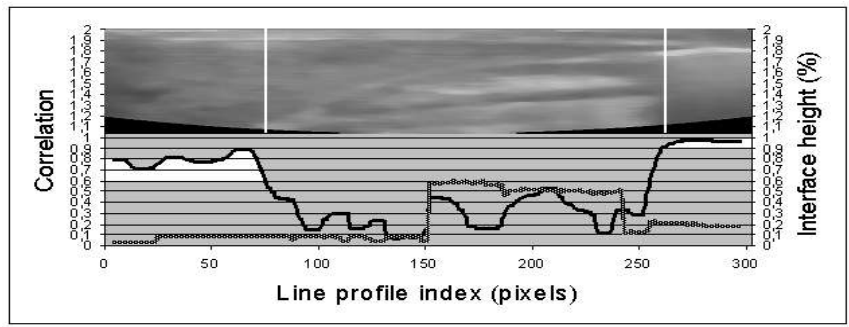

**Fig. 3.** Correlation profile (black curve) and height of shadow interface (grey curve) along the US fan. Left and right white areas correspond to shadow regions (high correlation, low height)

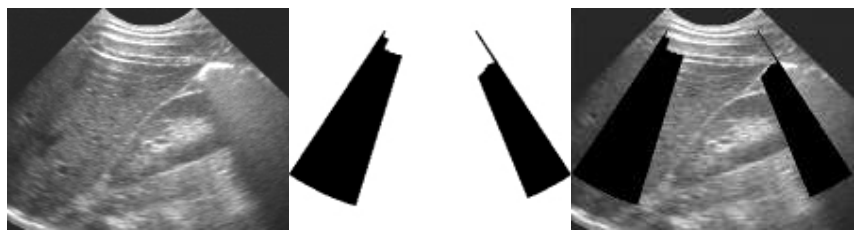

**Fig. 4.** US shadow removal. Left image shows the shadow profiles induced by the ribs (left) and the colon (right). Middle image shows the generated mask, superimposed on right image

## 3 Search Strategy: Powell-Brent Algorithm

PB algorithm [2][4] appears as a fairly efficient search strategy when the differentiation of the cost function is unknown [4]. Our implementation is based on [7]. We applied some changes since we found the method too slow and too sensitive to local minima, which are frequent in image registration.

### 3.1 Initial Attitude and Parameter Space

The minimization process consists in optimizing a 6D vector $(T_X,T_Y,T_Z,R_X,R_Y,R_Z)$. From the Arun-based Initial Attitude (IA) computation, we define the longitudinal axis of the CT kidney as the Z' axis of the new parameter space; this allows us to follow Maes' advice [5] to optimize the vector in the better-conditioned order $(T_{X'},T_{Y'},R_{Z'},T_{Z'},R_{X'},R_{Y'})$.

### 3.2 Search Interval

Contrary to fundamental PB method, whose Brent 1D search interval, for each DOF, is defined by a time-consuming automatic bracketing, we chose an arbitrary interval length of $2*RMS_{ARUN}$ (20 to 30mm) around the IA.

Since PB tends to reach the solution very fast (only the first iteration makes a significant step), it was decided, for computation time improvement, to reduce the interval length at each Powell iteration.

### 3.3 1D Initial search

As PB method tends to converge to local minima, and as the similarity between registered CT and US shows in most cases a non-ambiguous global minimum (fig. 6), we chose to perform before each Brent 1D optimization an initial search, with a step of 10% of the interval length, to find an approximate global minimum in that DOF. Then Brent method is applied around that minimum in a very restrained interval (fig. 5). This strategy does not significantly increase the computation time. Jenkinson [2] also proposed an initial search on the rotational DOF, previously to PB iterations, to make the solution "more reliable".

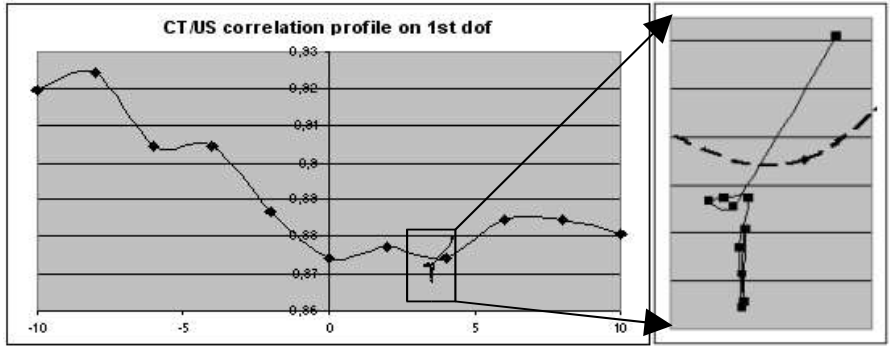

**Fig. 5.** CR profile along the search interval for $T_X$. The converging curve represents Brent iterations around the global minimum of the profile

## 4 Similarity Measure: Correlation Ratio

### 4.1 Definition

Let X be the base image (CT), and Y the match image (US). CR is a functional similarity measure [2][8][10] based on conditional moments. It measures the functional relationship between the match pixels and the base pixels. Roche [8] writes:

$$\eta(Y|X) = 1 - \frac{1}{N\sigma^2} \sum_i N_i \sigma_i^2 \qquad (0)$$

where N is the number of overlapped pixels, and σ their variance in Y.
where $N_i$ is the number of overlapped pixels in X of value i, $\sigma_i$ their variance in Y.

### 4.2 Motivation

Theoretically, the functional relationship between an intensity CT image and a gradient US image is not obvious, thus a statistical similarity measure like normalized mutual information (NMI) should be more appropriate. However, although US images best enhance the boundaries between organs, they also convey an information on the tissues, as each of them diffuses the US waves its own way, in both intensity and frequency. So CR is somehow justified, especially with the preprocessing we presented above. We also chose CR against NMI for several reasons:

- CR looks more continuous than NMI (fig. 6); few local minima appear, even when subsampling: it is more precise, discriminative and robust [10].
- CR algorithm complexity is $O(N_X)$, where $N_X$ is the number of overlapped grey levels in X, whereas NMI algorithm complexity is $O(N_X N_Y)$ [8].
- CR continuous profile is particularly adapted to PB search algorithm, since Brent minimization is based on the fitting of a parabola onto the function curve [7].

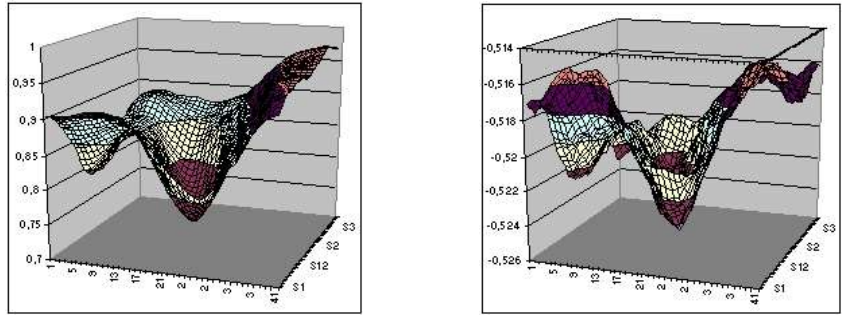

**Fig. 6.** CR (left) and NMI (right) profiles mapped along $T_X$ and $T_Y$ while registering 2 images. Note the continuous aspect of CR and the unambiguous minimum

### 4.3 2.5D Correlation Ratio

In the context of 2.5D US, the CR is computed as follows: for every rigid transform estimate, the CT volume is resliced in the plane of each US slice, using linear interpolation. Then the region of interest (ROI) defined in each US image is superimposed onto the corresponding CT slice. Finally, the CR is estimated on the set of pixels of the superimposed ROIs.

## 5 Preliminary Results and Discussion

### 5.1 Material and Methods

We worked on a single data set. The isotropic abdominal CT volume has 287x517x517 voxels, the voxel size being 0.6mm. Our experiments were carried out on a late acquisition CT, after iodine injection. Besides, for the use of our experiments, 100 US images (axial and longitudinal) of the right kidney were acquired with a localized calibrated echographic probe, through anterior access. Those images are then resized to match CT voxel size. In the final protocol, each registration will be carried out on only 5 slices of various orientations out of the whole set.

The accuracy study consists in performing 2 series of 10 registrations: on the one hand, by fixing the IA and changing the set of 5 US slices to match; on the other hand, by computing a new IA at each attempt, while keeping the same set of slices. Every IA is determined by the user from the choice of 4 points at the axial and lateral extremities of the kidney.

From our previous study [3] we have a "Bronze" Standard (BS) transform based on ICP matching, and segmented US and CT meshes. The quality of the solution is evaluated by the distance of the registered CT mesh to the BS CT mesh, and the difference between their gravity centers and their major axes inclinations.

### 5.2 Results

The registrations took on average 80s[1]. The algorithm appeared stable since a second matching pass kept the kidney in place. Table 1 shows matching statistics between the CT mesh transformed by our algorithm and the CT mesh obtained by the BS transform. *D(CT,BS)* (mm) shows the distance of the CT mesh to the BS mesh. Δ *Center* (mm) and Δ *Inclination* (°) are the differences between gravity centers and major axis inclinations. The choice of the criteria will be discussed later. Generally the IA sets the kidney at about 10mm in translation and 10° in inclination from the solution. For example, with IA n°1 we obtained: 12.3mm±2.6 ; 9.6mm ; 9.0°. With 1mm to 6mm error we can say that every registration has succeeded, except n°13 and n°17 whose position and orientation hardly improved.

**Table 1**. Matching statistics when the IA is constant and when the set of slices is fixed

| Constant IA | | | | Constant set of slices | | | |
|---|---|---|---|---|---|---|---|
| Nb | D(CT,BS) | Δ Cent. | Δ Inclin. | Nb | D(CT,GS) | Δ Cent. | Δ Inclin. |
| 1 | 5.6+/-2.1 | 2.7 | 4.1 | 11 | 3.7+/-1.3 | 1.3 | 4.3 |
| 2 | 6.8+/-2.0 | 3.2 | 6.7 | 12 | 3.2+/-0.8 | 2.9 | 1.5 |
| 3 | 5.2+/-1.5 | 3.2 | 1.3 | 13 | 8.8+/-2.5 | 5.7 | 10.5 |
| 4 | 6.2+/-1.8 | 4.0 | 5.3 | 14 | 5.7+/-2.0 | 4.0 | 5.3 |
| 5 | 5.8+/-1.8 | 2.3 | 9.5 | 15 | 6.7+/-2.6 | 5.1 | 8.2 |
| 6 | 6.0+/-2.0 | 4.3 | 4.7 | 16 | 6.0+/-1.3 | 5.8 | 3.2 |
| 7 | 4.1+/-1.1 | 2.7 | 5.5 | 17 | 11.9+/-4.8 | 8.1 | 6.4 |
| 8 | 5.1+/-1.1 | 3.4 | 5.2 | 18 | 6.5+/-2.1 | 5.0 | 7.6 |
| 9 | 5.1+/-1.9 | 3.0 | 5.4 | 19 | 6.2+/-1.6 | 5.6 | 6.0 |
| 10 | 3.7+/-0.9 | 2.1 | 2.9 | 20 | 5.4+/-0.9 | 4.1 | 4.3 |

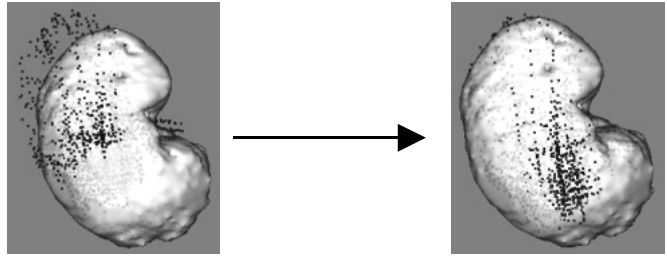

**Fig. 7.** Illustration of registration n°1 showing the US (black) and CT (white) meshes.

### 5.3 Discussion

We chose to compare homogeneous data, instead of computing the distance between the transformed CT mesh and the US cloud of points, as the bare information of an average distance between two sets of points does not appear relevant. To illustrate that, fig. 7 shows the result of matching n°1, which is visually very good. However, the distance of the US points to the IA-transformed CT mesh was 3.1mm±2.4 while the distance to the registered mesh was no smaller than 2.9mm±2.0. Furthermore, beside the quantitative evaluation of the solution, we also checked it visually, since the BS does not give an exact solution: in most cases the kidney position was acceptable for a further guiding procedure.

---
[1] Tests performed on a Pentium IV 1.7GHz

Although the results are satisfying in regard with accuracy, robustness and computation time, we must bear in mind that they highly depend on the choice of slices, the manual ROI, the manual IA and the search interval. The two failure cases were due to a bad IA. We expect to validate our method on other patient data sets in the following weeks. Anyway, the theoretical and practical difficulties in registering CT with US, particularly in terms of patient-dependent parameters settings, lead us to think that our method should be part of a higher-level algorithm that would also involve elastic registration, or the combination of rigid matching and deformable contour fitting, and certainly a minimal user intervention.